\documentclass[reprint,
 superscriptaddress,
 amsmath,amssymb,
 aps,
 prx,
 showkeys
]{revtex4-2}

\usepackage{amsmath}
\usepackage{amsthm}
\usepackage[colorlinks=true,urlcolor=blue,citecolor=blue,linkcolor=blue]{hyperref}
\usepackage[shortlabels]{enumitem}
\usepackage{amsfonts}
\usepackage{graphicx}
\usepackage{bbold}
\usepackage{color}
\usepackage{hyperref}
\usepackage{verbatim}
\usepackage{cases}
\usepackage{amsfonts}
\usepackage{amssymb}
\usepackage[]{mathrsfs}
\usepackage{xcolor}
\usepackage{epsfig}
\usepackage{bm}
\usepackage{setspace}
\usepackage{enumerate}
\usepackage{dcolumn}
\usepackage{bm}
\usepackage{enumitem}

\usepackage{graphicx}
\usepackage{dcolumn}
\usepackage{bm}
\usepackage[FIGTOPCAP,small,tight]{subfigure}
\usepackage{color}
\usepackage{amssymb}
\usepackage{amsmath}
\usepackage{graphicx}
\usepackage{array}
\usepackage{multirow}

\usepackage[normalem]{ulem}

\begin{document}

\preprint{APS/123-QED}

\title{Experimental realization of the classical Dicke model}

\author{Mario A. Quiroz-Ju\'{a}rez}
\thanks{\textcolor{blue}{These authors contributed equally.}}
\affiliation{Instituto de Ciencias Nucleares, Universidad Nacional Aut\'onoma de
M\'exico, Apartado Postal 70-543, 04510 CDMX, 
M\'exico}

\author{Jorge Ch\'avez-Carlos}
\thanks{\textcolor{blue}{These authors contributed equally.}}
\affiliation{Instituto de Ciencias F\'isicas, Universidad Nacional Aut\'onoma de
M\'exico, Av. Universidad s/n Col. Chamilpa, C.P. 62210 Cuernavaca, Morelos,
M\'exico}

\author{Jos\'e L. Arag\'on}
\affiliation{Centro de F\'isica Aplicada y Tecnolog\'ia Avanzada, Universidad Nacional Aut\'onoma de
M\'exico, Boulevard Juriquilla 3001, 76230 Quer\'etaro, Juriquilla, M\'exico}

\author{Jorge G. Hirsch}
\email{hirsch@nucleares.unam.mx}
\affiliation{Instituto de Ciencias Nucleares, Universidad Nacional Aut\'onoma de
M\'exico, Apartado Postal 70-543, 04510 CDMX, 
M\'exico}

\author{Roberto de J. Le\'on-Montiel}
\email{roberto.leon@nucleares.unam.mx}
\affiliation{Instituto de Ciencias Nucleares, Universidad Nacional Aut\'onoma de
M\'exico, Apartado Postal 70-543, 04510 CDMX, 
M\'exico}

\date{\today}

\begin{abstract}
We report the experimental implementation of the Dicke model in the semiclassical approximation, which describes a large number of two-level atoms interacting with a single-mode electromagnetic field in a perfectly reflecting cavity. This is managed by making use of two non-linearly coupled active, synthetic LC circuits, implemented by means of analog electrical components. The simplicity and versatility of our platform allows us not only to experimentally explore the coexistence of regular and chaotic trajectories in the Dicke model but also to directly observe the so-called ground-state and excited-state ``quantum'' phase transitions. In this analysis, the trajectories in phase space, Lyapunov exponents and the recently introduced Out-of-Time-Order-Correlator (OTOC) are used to identify the different operating regimes of our electronic device. Exhaustive numerical simulations are performed to show the quantitative and qualitative agreement between theory and experiment.

\end{abstract}

\pacs{Valid PACS appear here}
\maketitle

\section{Introduction}
The quantum dynamics of isolated many body systems can reach states of equilibrium, in some cases allowing for a thermal description, when they are in a chaotic regime \cite{Reimann2008,Short2011,Short2012,Zangara2013,HeSantos2013,Gogolin2016,Borgonovi2016,Alessio2016,Reimann2016,Reimann2018,Dymarsky2018}.
While equilibrium is associated with small fluctuations around an average value of some observables, whose relative ratio becomes smaller as the size of the system is increased, thermalization refers to the agreement of these average values with those obtained by means of statistical mechanics. Determining under which conditions a given quantum system equilibrates is both relevant and challenging. Interestingly, chaos plays a fundamental role in the equilibration of a quantum system under unitary dynamics \cite{Altland2012NJP,Altland2012PRL}. It makes particularly important the study of quantum systems whose classical counterparts are chaotic \cite{Berry1977,Berry1981,Bohigas1984}, a very common situation as the classical dynamics of many body systems in interaction is non-linear, non-integrable, chaotic, as they have less integrals of motion than degrees of freedom.

Originally introduced as the quantum description of collective behavior of two-level atoms interacting with a single electromagnetic mode of a cavity, the Dicke model \cite{dicke,Garraway2011} has found application in the study of equilibrium and thermalization of isolated quantum many-body systems, as well as their classical-quantum correspondence \cite{Schneble2003,Blais2004,Scheibner2007,Fink2009,Baumann2010,Nagy2010,Niemczyk2010,Casanova2010,Baumann2011,Mezzacapo2014}. Recent experiments with ion traps have shed new light on the system \cite{Cohn2018,Safavi2018}, and has opened the possibility of observing quantum indicators of chaos \cite{Lewis2019} in the Dicke model.

The Dicke model has been a subject of intensive research for many years. In particular, it has shown to exhibit quantum phase transitions in the ground \cite{Emary2003PRL,Emary2003,Castanos2009} and the excited states \cite{Cejnar2006,Perez2011,Brandes2013,Bastarrachea2014pra1,Blas2018}. Indications of the presence of quantum chaos in the fluctuation of its energy spectra were reported in \cite{Emary2003} and studied in detail in \cite{Bastarrachea2014pra2}. The presence of regular and chaotic regions has been analyzed qualitatively employing Poincar\'e sections for the classical analysis and Peres lattices in the quantum case \cite{Bastarrachea2015}, followed by a detailed quantitative analysis of the classical phase space and the parameter space performed by means of Lyapunov exponents \cite{chavez2016}. Along these lines, it has been proven that there exists a clear correlation between the Lyapunov exponents and the participation ratio of the coherent states in the Dicke Hamiltonian eigenbasis \cite{Bastarrachea2016}. Furthermore, the Out-of-Time-Order-Correlator (OTOC) has been shown to grow at short times with a ratio close to the Lyapunov exponent \cite{chavez2019}.

Surprisingly, an aspect which has never before been shown is a physical realization of the  classical Dicke Hamiltonian. In this work, we present its first implementation. This is done by making use of two synthetic, non-linearly coupled electrical LC (where L stands for inductance and C for capacitance) oscillators, implemented by means of analog electrical components. Our platform is thus used to observe the dynamics of the classical Dicke model and monitor important features, such as the ground-state and excited-state phase transitions, the coexistence of periodic and chaotic trajectories, and the OTOC dynamics. Because of its simplicity and versatility, our platform constitutes an excellent test bed for exploring the richness of non-linear systems in terms of chaos and regularity.

The paper is structured as follows. In Sec. II, we present the semiclassical approximation of the Dicke model and the Hamiltonian describing its classical dynamics. In Sec. III, we show how the classical Dicke Hamiltonian can be expressed in terms of non-linearly coupled harmonic oscillators, specifically, LC electrical oscillators. In Sec. IV, we describe the dynamical evolution of the system and discuss some representative phenomena of the classical Dicke model and finally, in Sec. V, we present our conclusions.

\section{Semiclassical Dicke Hamiltonian approximation}

In this section, we briefly describe the semiclassical approximation of the Dicke model and its corresponding Hamiltonian. We start by considering the quantum Hamiltonian for $N$ identical two-level atoms under the action of a single-mode quantized radiation field (Dicke model) \cite{dicke}, which writes ($\hbar=1$)

\begin{equation}
H_D=\omega \hat{a}^{\dagger}\hat{a}+\omega_0 \hat{J}_z+\frac{\gamma}{\sqrt{N}}\left(\hat{J}_+ +\hat{J}_-\right)\left(\hat{a}^{\dagger} +\hat{a}\right),
\label{eq:hamiltonian}
\end{equation}
\\
where $\omega$ is the frequency of the monochromatic quantized radiation field and $\omega_0$ is the excitation energy of a set of two-level atoms. $\hat{a}$ and $\hat{a}^{\dagger}$ represent the one-mode annihilation and creation photon operators, respectively. $\hat{J}_z$, $\hat{J}_+$ and $\hat{J}_-$ are collective atomic pseudo-spin operators which obey the \textit{SU(2)} algebra. Specifically, $\hat{J}_z$ is the atomic relative population operator and $\hat{J}_{\pm}$ are the atomic transition operators. If $j(j+1)$ is an eigenvalue of angular momentum operator $\hat{J}^2$ then $j=N/2$ defines the symmetric atomic subspace, including the ground state. Finally, $\gamma$ is the matter-light coupling strength, which depends on the atomic dipolar moment. When the coupling strength reaches the critical value $\gamma_c=\sqrt{\omega \omega_0}/2$, the ground state of the system goes from normal to a super-radiant phase. This behavior is the most representative phenomena of the Dicke Hamiltonian in the thermodynamic limit: the Quantum Phase Transition (QPT) \cite{Emary2003}. Another important feature of the Dicke Hamiltonian is the presence of an Excited State Quantum Phase Transition (ESQPT) manifested by simultaneous singularities in the eigenvalue spectrum,  order parameters, and wave function properties \cite{stransky2014, Perez2011, caprio2008}.

By employing Glauber coherent states given by $\left|\alpha\right>=e^{-\left|\alpha\right|^2/2}e^{\alpha \hat{a}^{\dagger}}\left|0\right>$ and Bloch coherent states described by $\left|z\right>=1/(1+\left|z\right|^2)^{-j}e^{z\hat{J}_+}\left|j,-j\right>$, one can find an effective classical Hamiltonian by taking the expectation value of the Hamiltonian operator in the coherent state product \cite{deaquilar1992}. The classical Hamiltonian per particle, $\left<\alpha,z\right|H_D\left|\alpha,z\right>/j$, in terms of the canonical variables $(p,q)$ and
$(P,Q)$, reads \cite{Bastarrachea2016,chavez2016},

\begin{equation}
H_{cl}=\frac{\omega}{2}(p^2+q^2)+\frac{\omega_0}{2}(P^2+Q^2)+\gamma q Q\sqrt{4-P^2-Q^2}-\omega_0.
\label{eq:classical}
\end{equation}
where $(p,q)$ and
$(P,Q)$ are pairs of real conjugate coordinates for the photonic and the atomic sector, respectively.
They are given in terms of the coherent state parameters as
\begin{eqnarray}
\alpha=\sqrt{\frac{j}{2}}(q+ip), \\
z=\frac{Q-iP}{\sqrt{4 -(Q^2+P^2)}} .
\end{eqnarray}

The dynamical properties of the Hamiltonian (\ref{eq:classical}) can be described by the temporal evolution of the canonical variables $\mathbb{X}=(p,q,P,Q)$, whose equations of motion are \cite{chavez2016}

\begin{equation}
\begin{aligned}
\dot{p}&=-\gamma Q\sqrt{4-P^2-Q^2}-q\omega,\\
\dot{q}&=p\omega,\\
\dot{P}&=\frac{\gamma q Q^2}{\sqrt{4-P^2-Q^2}}-\gamma q \sqrt{4-P^2-Q^2}-Q\omega_0,\\
\dot{Q}&=P\omega_0-\frac{\gamma PqQ}{\sqrt{4-P^2-Q^2}}.
\end{aligned}
\label{eq:motionpq}
\end{equation}
\\
Note that the field variables $(p,q)$ are unbound, while the atomic variables $(P,Q)$ must satisfy the condition
$P^2+Q^2 \leq 4$, reflecting the fact that the quantum angular momentum space employed to describe the collective atomic excitations is finite, with dimension $2j+1$. In the classical dynamics governed by Hamiltonian  (\ref{eq:classical})  there is no reference to $j$, this inequality is a memory of it in the classical realm, with the point $P=Q=0$ associated with the south pole of the Bloch sphere, and the circumference defined by $P^2+Q^2 = 4$ with the north pole.

The equilibrium points can be easily determined by equating the dynamical equations (\ref{eq:motionpq}) to zero,  $\mathbf{F}(\mathbb{X})= \overline {0}$,  and simultaneously solving the set of algebraic equations. In doing so, we obtain the following critical points:

\begin{eqnarray}\label{eq:equilibrium}
\mathbb{X_0}&=&(0,0,0,0), \nonumber \\\mathbb{X_\pm}&=&\left(0,\pm\frac{2 \gamma^2}{\omega}\sqrt{1-\frac{\gamma_c^4}{\gamma^4}},0,\mp \sqrt{2\left(1-\frac{\gamma_c^2}{\gamma^2}\right)}\right),
\end{eqnarray}
\\
Note that if $\gamma>\gamma_c$, the system has three real equilibrium points, and when $\gamma\le\gamma_c$, $\mathbb{X_0}$ is the only real equilibrium point. This result is consistent with the so-called saddle-node bifurcation, characterized by the annihilation of two points of equilibrium in a dynamical system \cite{kuznetsov2013}. Next, we consider the dynamics of the system in the neighborhood of $\mathbb{X}_0$. The Jacobian matrix $\mathbb{J}=D_\mathbb{X}\mathbf{F}(\mathbb{X})$  evaluated at this point is given by,

\begin{equation}
\mathbb{J}^\mathbb{X_0}=\left(
\begin{array}{cccc}
 0 & -\omega  & 0 & -2 \gamma  \\
 \omega  & 0 & 0 & 0 \\
 0 & -2 \gamma  & 0 & -\omega_0  \\
 0 & 0 & \omega_0  & 0 \\
\end{array}
\right),
\label{eq:jacobian}
\end{equation}
\\
and its eigenvalues, in the resonance case $\omega=\omega_0$, are described by $\lambda(\mathbb{J}^\mathbb{X_0})=\left\{\mp 2\gamma_c \sqrt{\frac{\gamma}{\gamma_c}-1},\mp2i\gamma_c \sqrt{\frac{\gamma}{\gamma_c}+1}\right\}$. It follows from the above that Eq. (\ref{eq:jacobian}) has two real and two purely complex eigenvalues (saddle point) when the interaction strength $\gamma>\gamma_c$ and become purely imaginary (center point) when the interaction strength is smaller, i.e., $\gamma<\gamma_c$. This stability change signals the so-called Andronov-Hopf bifurcation, in which an equilibrium switches its stability via a pair of purely imaginary eigenvalues \cite{andronov1971, kuznetsov2013}.
By making a similar analysis in the equilibrium points $\mathbb{X}_{\pm}$, one can find that they correspond to center points. Notably, for $\gamma=\gamma_c$ and $E=-1$, both center points collide with the saddle point, giving rise to a phenomenon that resembles a homoclinic bifurcation \cite{Bastarrachea2014pra1,Bastarrachea2014pra2,stransky2014,kuznetsov2013}.

At the transition point $\gamma=\gamma_c$ and $E=-1$, $\mathbb{X}_0$ undergoes a Bogdanov-Takens (BT) bifurcation \cite{arnold2012,kuznetsov2013}, which meets the three codimension-one bifurcations: saddle-node bifurcation, Andronov-Hopf bifurcation and homoclinic bifurcation. Additionally, the Jacobian matrix around the equilibrium point given by Eq. (\ref{eq:jacobian}) presents a zero eigenvalue of multiplicity two.

\section{Dicke model -- LC circuit analogy}

In recent years, electronic analogies have been successfully used to understand key concepts and/or explore new focuses of complex quantum phenomena. In particular, they have allowed for building experimental systems that offer certain implementation advantages---such as cost, complexity, and measuring protocols---over the original setups \cite{leon2015, quiroz2017, leon2018}. It is because of these features that electronic circuits have arisen as a powerful tool for exploring observables that obey the same mathematical description as those of the original quantum or classical systems.

Interestingly, the Hamiltonian (\ref{eq:classical}) can be understood as two harmonic oscillators, one of them representing the set of two-level atoms $(P,Q)$ and the other the radiation field $(p,q)$, non-linearly coupled through its atomic dipolar moment. This allows us to map the original Hamiltonian (\ref{eq:classical}) to an electrical version of the harmonic oscillator, that is, LC circuits which are non-linearly coupled to each other, with L standing for inductance, and C for capacitance. We can thus rewrite the Hamiltonian (\ref{eq:classical}) in terms of the electrical variables by mapping $(p,q,P,Q)\rightarrow(\tilde{V}_{C_1},\tilde{I}_{L_1},\tilde{V}_{C_2},\tilde{I}_{L_2})$ with $\tilde{V}_{C_1}=\sqrt{\omega C_1}V_{C_1}$, $\tilde{I}_{L_1}=\sqrt{\omega L_1}I_{L_1}$, $\tilde{V}_{C_2}=\sqrt{\omega_0 C_2}V_{C_2}$ and $\tilde{I}_{L_2}=\sqrt{\omega_0 L_2}I_{L_2}$, where $\tilde{V}_{C_j}$ is the normalized voltage in the $j$th capacitor, and $\tilde{I}_{L_j}$ is the normalized current passing through of the $j$th inductor, with $j=1,2$. The system described by the new electrical variables can then be experimentally implemented using active electrical networks of capacitors, resistors, operational amplifiers, and analog multipliers (see Appendix A for details). In our experimental setup, the parameters and the initial conditions are established by controllable voltage levels, in other words, the values of $\omega$, $\omega_0$, $\gamma$, $\tilde{V}_{C_1}(0)$,$\tilde{V}_{C_2}(0)$, $\tilde{I}_{L_1}(0)$ and $\tilde{I}_{L_2}(0)$ are directly mapped to
electrical potential differences provided by high-resolution digital-to-analog converters.

To explore the dynamical evolution of the classical model under different conditions, we follow the voltages of each oscillator (voltage in capacitors and current through inductors) corresponding to the canonical variables $(p,q,P,Q)$, in order to collect information about the energy of the system. We extract time series of 300 s sampled at 15 ms using an oscilloscope Tektronix TBS200 (impedance 1M$\Omega$). Electronic components were mounted and soldered on a printed circuit board (15x15 $c$m) to avoid faulty contacts and poor stability.

\begin{figure*}[t]
\centering
\includegraphics[width=18 cm]{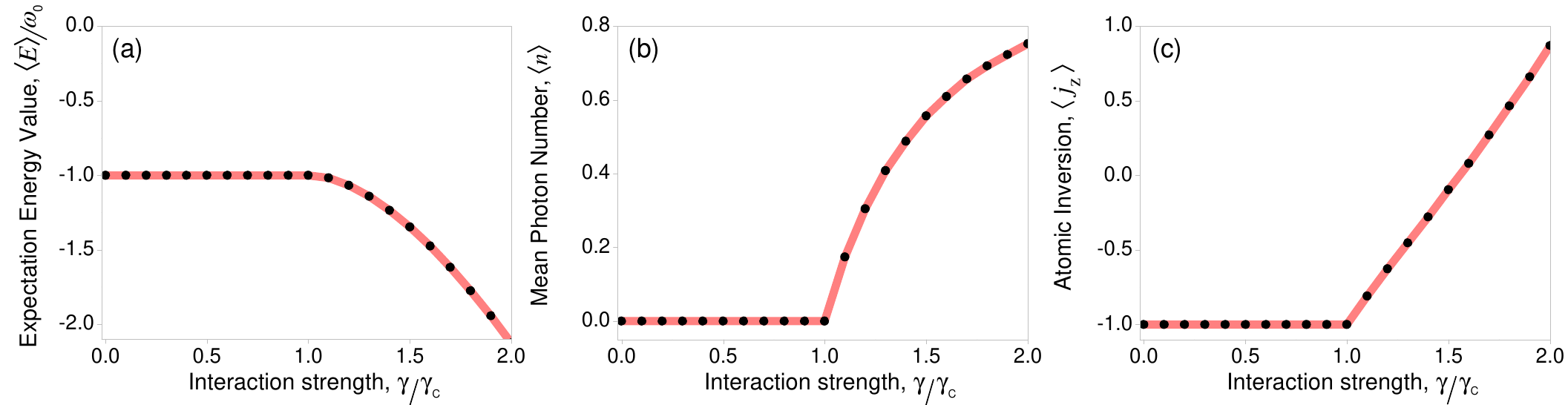}
\caption{(a) Expectation energy value of the ground state, (b) Classical analogous of the mean photon number $\left\langle n \right\rangle$ and (c) Classical analogous of the atomic inversion $\left\langle j_z \right\rangle$ as a function of the interaction strength, for the classical Dicke model. The black points correspond to experimental data, whereas the red lines denote theoretical results. The parameters are set to $\omega=1$ and $\omega_0=1$ s$^{-1}$. The standard deviation of the experimental results are inside the black points.}
\label{fig:ground}
\end{figure*}

\section{Results}

One of the merits of the proposed setup is the possibility of exploring many remarkable features of the classical Dicke model by easily accessible parameter configurations. To illustrate the functionality of our experimental setup, we carry out measurements of some representative phenomena of the Hamiltonian (\ref{eq:classical}). In what follows, we compare experimental and theoretical results for the ground-state and excited-state phase transition, and the dynamics of different regular and chaos regions, all in the resonant case ($\omega=\omega_0$). Additionally, we calculate the OTOC for an ensemble of random initial conditions.

\subsection{Ground-State Phase Transition (GSPT)}

In the thermodynamic limit, i.e. when the number of atoms $N$ goes to infinity, the Dicke model presents a quantum phase transition (whenever the interaction strength reaches a critical value) characterized by a discontinuity in the second derivative of the minimal energy \cite{nahmad2013}. This can be observed by following the ground state energy of our experimental setup and comparing it with the semiclassical ground state energy $E_0(\gamma)$ given by,

\begin{equation}
E_0(\gamma)=\left\{\begin{matrix}-\omega_0 &  & $for$  \quad \gamma \le \gamma_c \\  &  &  \\ -\frac{\omega_0}{2}\left(\frac{\gamma_c^2}{\gamma^2}+\frac{\gamma^2}{\gamma_c^2} \right) &  & $for$  \quad \gamma > \gamma_c \end{matrix}\right.
\label{eq:groundstate}
\end{equation}

The experimental average energy, along a given trajectory $\left< E \right>$, was calculated using measured data of our experimental setup through the following expression

\begin{equation}
\left< E \right> = {\frac{1}{t}\int_{0}^{t}{H_{cl}}d\tau},
\end{equation}
\\
where $t$ corresponds to the observation time. Figure \ref{fig:ground}(a) shows the expectation energy value of the ground state, in the classical Dicke model, as a function of the interaction strength, for both (black points) experimental measurements and (red line) analytical expression (\ref{eq:groundstate}). Qualitative and quantitative agreements between both results can clearly be observed.

We can further use the measured data to plot the classical analogous of the mean photon number $\left\langle n \right\rangle$ and the atomic inversion  $\left\langle j_z \right\rangle$ as a function of the interaction strength. These are given by $(p^2+q^2)/2$ and $(P^2+Q^2)/2$, respectively [see Figs. \ref{fig:ground}(b) and (c) ]. Note that the ground state undergoes a sudden change in its properties, going from an unexcited normal phase $(\gamma\le\gamma_c)$ with no photons and no excited atoms to a symmetry-broken super-radiant phase $(\gamma>\gamma_c)$, in which both the mean number of photons and the number of atoms in the excited state become comparable to the total number of atoms in the system, in other words, the field and atomic collection acquire macroscopic occupations. This transition is an example of a quantum collective behavior and has a close connection with entanglement and quantum chaos \cite{Emary2003}.

\begin{figure*}[t!]
\centering
\includegraphics[width=18cm]{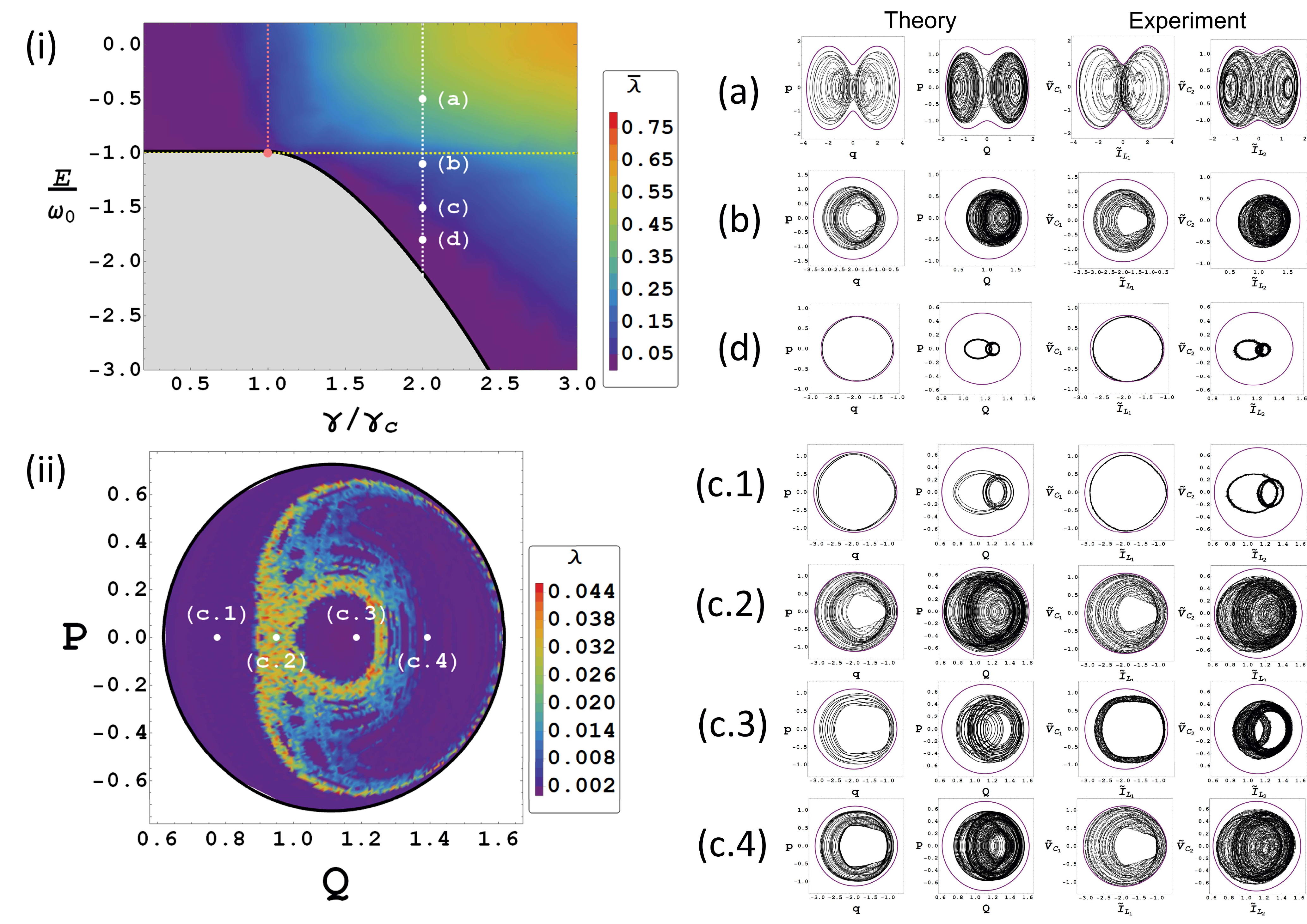}
\caption{($i$) Map of the average Lyapunov exponents as a function of the energy $E$ and the coupling strength $\gamma/\gamma_c$. The thick black line follows the ground-state energy and the red dot denotes the critical point $\gamma_c$. The white
dots indicate four different energies for a fixed coupling $\gamma=2\gamma_c$:
$(a)$ $E=-0.5$, $(b)$ $E=-1.1$, $(c)$ $E=-1.5$ and $(d)$ $E=-1.8$. $(ii)$ Map of chaoticity in the plane $(Q,P)$, for $p=0$ and $E = -1.5$ (mixed chaos zone). Four different initial conditions are  marked with white points into the map of chaoticity.  Right hand side shows the projections of the trajectories in the planes $(q,p)$ and $(Q,P)$ for the numerical simulations, and in the planes ($\tilde{I}_{L_1},\tilde{V}_{C_1}$) and ($\tilde{I}_{L_2},\tilde{V}_{C_2}$)
for the experimental results, corresponding to the initial points $(p,q,P,Q)$:
$(a)$ (0,-0.13372,  0, 1.22474),
$(b)$ (0,-0.50471,  0, 1.22474),
$(c.1)$ (0,-0.76496,  0, 0.7746),
$(c.2)$ (0,-0.72693, 0, 0.94868),
$(c.3)$ (0,-0.79433, 0, 1.18322),
$(c.4)$ (0,-1,  0,1.41421) and
$(d)$ (0,-1.0996,  0, 1).
The purple lines in the phase planes denote the available evolution space. The time evolution runs up to 300s, and we set  $\omega=1$ s$^{-1}$ and $\omega_0=1$ s$^{-1}$.
}
\label{fig:mapLyapunov}
\end{figure*}

\subsection{Regularity and chaos}
In this subsection, we explore the presence of regularity and chaos for different energies. By employing the Lyapunov exponents and phase space, we determine the nature of the trajectories for both theoretical and experimental results.
Figure \ref{fig:mapLyapunov}(i) shows the map of the average Lyapunov exponents \cite{liapunov, oseledets1968, benettin1980} as a function of the energy $E$ and the coupling strength $\gamma/\gamma_c$. For each pair ($E$,$\gamma/\gamma_c$) the
average Lyapunov exponent $\bar{\lambda}$ is calculated by simultaneously solving the dynamical equations (\ref{eq:motionpq}) and those in the corresponding tangent space
for a thousand of initial conditions in the restricted phase space defined by $E$ and $p=0$ \cite{chavez2016}. The thick black line denotes the ground-state energy and the red dot the critical point $\gamma_c=\sqrt{\omega \omega_0}/2$, which marks the transition between the normal and super-radiant phase. Note that the range of possible energies $E$ is lower bounded by the ground state [gray zone in Fig. \ref{fig:mapLyapunov}(i)]. There is no upper bound in the energy because the number of bosons in the field is not limited.

By looking at the map of the average Lyapunov exponents, one can appreciate the wide rich phenomenology in terms of chaos and periodicity. We have selected four different energies, in resonance and for the interaction strength $\gamma=2\gamma_c$. Each value is indicated on the phase map by white points: $(a)$ $E=-0.5$, $(b)$ $E=-1.1$, $(c)$ $E=-1.5$ and $(d)$ $E=-1.8$.

Figure \ref{fig:mapLyapunov}(ii) displays a detailed map of chaoticity in the
plane $(Q,P)$, for $p=0$ and $E = -1.5$, a mixed zone where regular and chaotic orbits coexist. Four different initial conditions $(p,q,P,Q)$ are selected and marked with white dots, as representatives of the different dynamics.

\begin{figure*}[t!]
\centering
\includegraphics[width=18 cm]{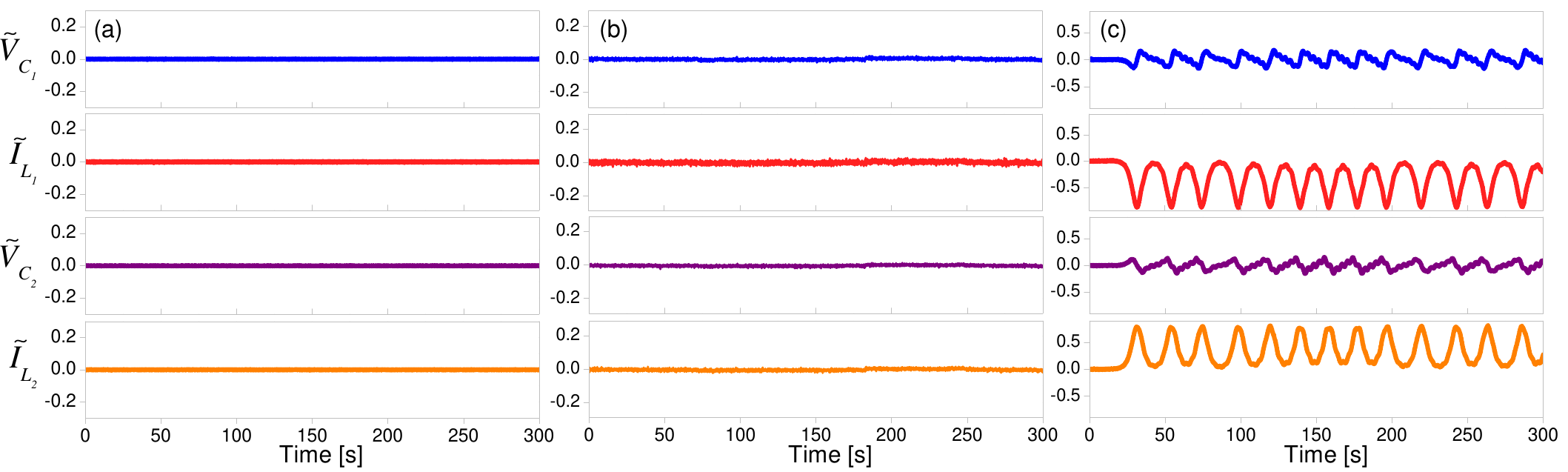}
\caption{Temporal evolutions of the electrical coordinates $(\tilde{V}_{C_1},\tilde{I}_{L_1},\tilde{V}_{C_2},\tilde{I}_{L_2})$, obtained from experimental data for three cases: $(a)$ Normal phase ($\gamma=0.9\gamma_c$) with $E_0=-0.99997\omega_0$, $(b)$ QPT ($\gamma=\gamma_c$) with $E_0=-0.999984\omega_0$ and $(c)$ Super-radiant phase ($\gamma=1.1\gamma_c$) with $E_0=-0.999865\omega_0$.}
\label{fig:esqpt}
\end{figure*}

The projections of the trajectories in the planes $(q,p)$ and $(Q,P)$ for the numerical simulations, and in the planes ($\tilde{I}_{L_1},\tilde{V}_{C_1}$) and ($\tilde{I}_{L_2},\tilde{V}_{C_2}$) for the experimental results, corresponding to the initial points (a),(b),(c.1),(c.2),(c.3),(c.4) and (d), are shown on the right hand side of Fig. \ref{fig:mapLyapunov}. For energies larger than $E=-1$, the dynamics are fully chaotic in an ergodic sense. At these energies, strong chaos characterized by Lyapunov exponents beyond $\lambda=0.15$ can be found. Note that $\lambda$ refers to the Lyapunov exponent for each initial condition shown in Fig. \ref{fig:mapLyapunov}(ii); whereas $\bar{\lambda}$ in Fig. \ref{fig:mapLyapunov}(i) refers to the average of all initial conditions. Figure \ref{fig:mapLyapunov}(a) shows a dense phase plane for $E=-0.5$, indicating the presence of chaotic dynamic. For energies close to $E=-1$, chaotic trajectories predominate. As Figure \ref{fig:mapLyapunov}$(b)$ shows, the phase plane exhibits a so-called strange attractor for $E=-1.1$. When the energy decreases up to $E=-1.5$, the complexity and non-linearity of the system acquire relevance in the dynamics, thus giving rise to a mixed zone where chaos and regularity coexist. Figures \ref{fig:mapLyapunov}(c.1)-(c.4) display phase planes for the sub-cases marked into the map of chaoticity [Fig. \ref{fig:mapLyapunov}(ii)]. Note that Figs. \ref{fig:mapLyapunov}(c.1) and (c.3) show regular dynamics; whereas the planes in Figs. \ref{fig:mapLyapunov}(c.2) and (c.4) present chaotic behavior. Finally, when $E=-1.8$, a limit cycle is formed, see Fig. \ref{fig:mapLyapunov}$(a)$.

To verify the nature of the experimental trajectories, we computed the largest Lyapunov exponents $(\lambda_{EXP})$ by applying the method described in Ref \cite{rosenstein1993} and Appendix B. For comparison purposes, we have also calculated theoretical  Lyapunov exponents $(\lambda_{THE})$. The results have been summarized in Table \ref{tab:data}. From our previous studies \cite{chavez2016}, the values $\lambda < 0.004$ are considered null, because they are below the numerical error.

\begin{table}[t!]
\centering
\caption{Lyapunov Exponents}
\label{tab:data}
\resizebox*{4cm}{!}{
{\setlength{\arrayrulewidth}{0.2mm}
\begin{tabular}{cccc}
\hline
 &$E$&$\lambda_{EXP}$ &$\lambda_{THE}$
\\
\hline
$(a) $& -0.5 & 0.289 & 0.295
 \\
$(b)$& -1.1 & 0.118 &0.124
\\
$(c.1)$& -1.5 & 0.000 & 0.000
  \\
$(c.2)$& -1.5 & 0.009 & 0.011
  \\
$(c.3)$& -1.5 & 0.001 & 0.001
  \\
$(c.4)$& -1.5 & 0.011 & 0.012
  \\
$(d)$& -1.8 & 0.000 & 0.000
 \\
\hline
\end{tabular}}}
\end{table}

\subsection{Excited-State Phase Transition (ESPT)}

Similarly to the transition experienced by the ground state, the classical Dicke model exhibits a excited state quantum phase transition (ESQPT), which occurs along the energy spectrum for fixed parameters of the Hamiltonian. This phenomenon denotes a  singularity in the spectrum of quantized energy levels of the system which affects both density and dynamics levels \cite{stransky2015}, in fact, the ESQPT can be identified by studying the density of states in the semiclassical limit. Interestingly, such singularity leads to dramatic dynamical consequences \cite{caprio2008}.

Note that the gap between the ground state and critical energy $E=-1$ [dotted white horizontal line in Fig. \ref{fig:mapLyapunov}(i)] vanishes as the interaction strength decreases. Interestingly, at $\gamma=\gamma_c$ and $E=-1$, two of three critical points become complex and $\mathbb{X}_0$ point is the only real one; in other words, the system goes from three equilibrium points in the super-radiant phase to only one in the normal phase. Importantly, the critical points provide useful information to follow the behavior of the energy minimum, and their stability changes give rise to dramatic alterations of the accessible phase plane. Such alterations unveil the presence of an ESQPT. In the Dicke model, for energies in the sub-critical case $(E<-1)$, the available phase plane is formed by two disconnected lobes. At the critical energy value $E=-1$, both lobes collide at the point $(0,0,0,0)$. Finally, for energies in the super-critical case $(E>-1)$, these lobes merge, allowing the trajectories freely evolve in both lobes.

In the Dicke model, ESQPT is identified by studying the stability conditions of $\mathbb{X_0}$ point as a function of the interaction strength. This point goes from a center point in the normal phase to a hyperbolic point in the super-radiant phase. It is clear that the initial condition $(p,q,P,Q)=(0,0,0,0)$ leads to a stationary solution irrespective of $\gamma$ value, due to the fact that it is located just at the equilibrium. When the interaction strength is smaller than its critical value, for any arbitrarily small perturbation $\delta\mathbb{X}=(\delta p,\delta q,\delta P,\delta Q)$  around $\mathbb{X_0}$ (center point), the system presents stationary solutions.  Beyond the critical interaction strength, $\mathbb{X_0}$ becomes a saddle point, which for any perturbed state $\delta\mathbb{X}$ escapes out of the equilibrium, exhibiting irregular oscillatory solutions.

Figure \ref{fig:esqpt} shows the temporal evolution of the electrical coordinates $(\tilde{V}_{C_1},\tilde{I}_{L_1},\tilde{V}_{C_2},\tilde{I}_{L_2})$ measured for arbitrarily small perturbed initial conditions, considering three different interaction strengths: (a) $\gamma=0.9\gamma_c$ (normal phase), (b) $\gamma=\gamma_c$ (QPT) and (c) $\gamma=1.1 \gamma_c$ (super-radiant). As predicted, for couplings $\gamma$ before $\gamma_c$, all solutions are stationary, that is, the trajectories converge to the equilibrium point $\mathbb{X_0}$, as depicted in Fig. \ref{fig:esqpt}(a). For $\gamma=\gamma_c$, exactly at the QPT, the solutions are yet stationary but the time series begin to exhibit slight alterations as shown in Fig. \ref{fig:esqpt}(b). In striking contrast, when the couplings becomes larger than $\gamma_c$, any small perturbation leads to irregular trajectories, specifically chaotic dynamics [see Fig. \ref{fig:esqpt}(c)]. There, the initial condition does not converge to the equilibrium point, thus making evident the instability of the hyperbolic point. We would like to stress that the electronic noise associated with the electrical components in our experimental setup, introduces an unavoidable perturbation in the initial state. In this way, it is not necessary to physically prepare a perturbed initial condition in the system, the stability or instability of the critical point is naturally unveiled by the electronic fluctuations of the device.

 \begin{figure}[h]
\centering
\includegraphics[width=8 cm]{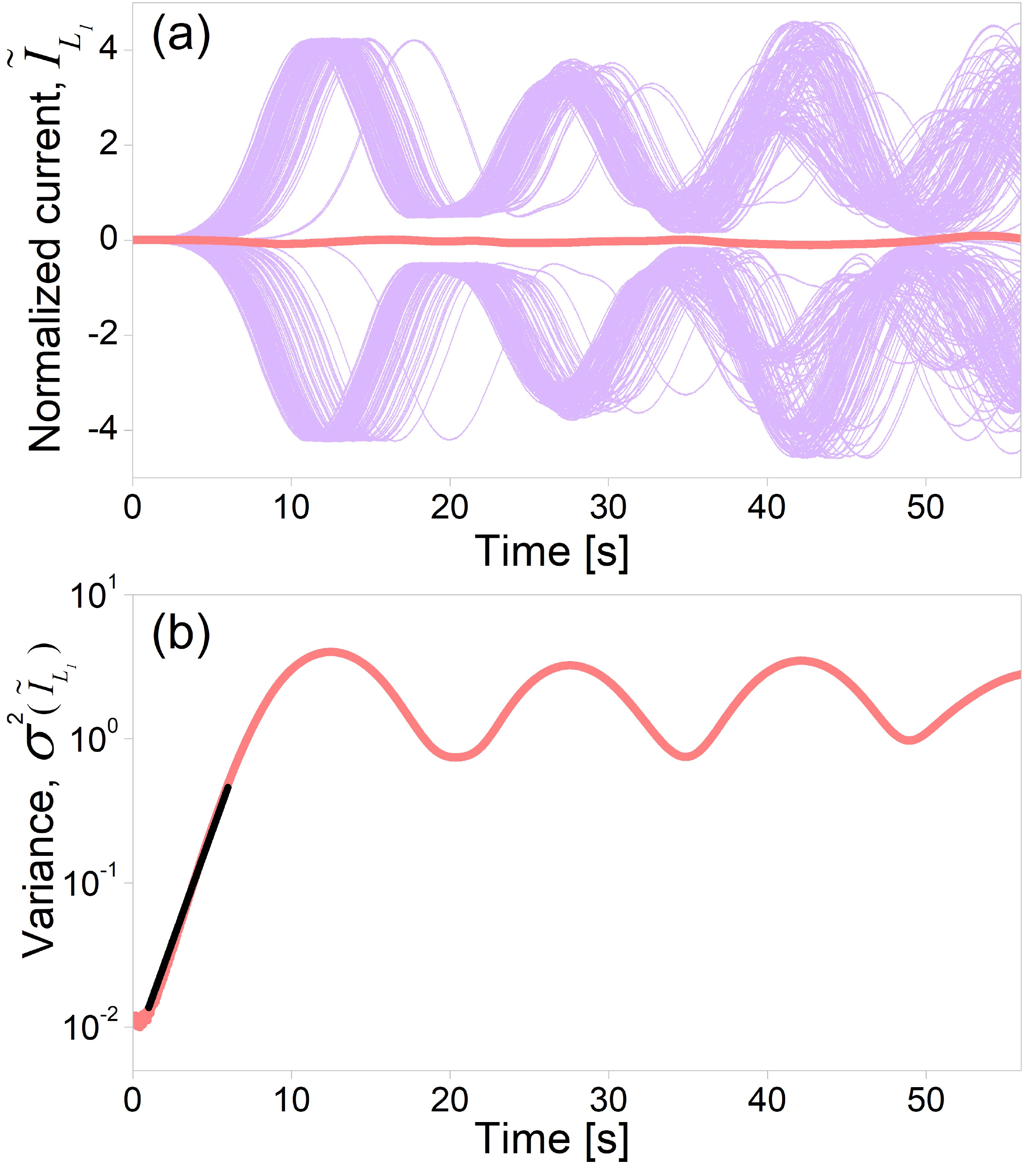}
\caption{(a) Experimentally measured time series for two hundred fifty-six randomly distributed initial conditions (purple lines) for the normalized current passing through one of the inductors $(\tilde{I}_{L_1})$. Red line corresponds to the mean current as a function of time. (b) Variance of the normalized current $\sigma^2(\tilde{I}_{L_1})$ as a function of time. Black line denotes the associated fit for $\sigma^2(\tilde{I}_{L_1})$ in the short-time behavior. The time evolution runs up to 60 s, and we set $\omega=0.5$ s$^{-1}$, $\omega_0=0.7$ s$^{-1}$ and $\gamma=0.66$.}
\label{fig:otoc}
\end{figure}

\subsection{Out-Of-Time-Order Correlator (OTOC)}

Recently, it has been demonstrated that the exponential growth rate of the out-of-time-order correlator (OTOC) is comparable with the classical Lyapunov exponent at short-time. Therefore, OTOC is now used as a measure of quantum chaos \cite{Lewis2019,chavez2019}. In this subsection, we compute the OTOC for an ensemble of two hundred fifty-six initial conditions in the super-radiant phase (chaotic energy region) by employing experimental data.

Figure \ref{fig:otoc}(a) shows the temporal evolution of the normalized current in one of the inductors $(\tilde{I}_{L_1})$, considering two hundred fifty-six randomly distributed initial conditions, which were acquired from our experimental setup. The red solid line represents $\left< \tilde{I}_{L_1} \right>$ along the time. Random initial conditions are generated by the electronic noise, which creates fluctuations in the voltages, making the system explore a region in phase space whose size is proportional to these fluctuations. This region is a ``classical uncertainty region'' in the sense that the values of the observables at each time cannot be determined with higher resolution. It has a parallel in the semiclassical description of the quantum evolution known as Truncated Wigner Approximation (TWA) \cite{steel1998, polkovnikov2010, schachenmayer2015many}. The TWA allows us to compute the quantum dynamics, up to the Ehrenfest time,  using the classical equations of motion, but averaging the observable over various different initial conditions. The random sampling reproduces the quantum fluctuations of a quantum initial state, and are usually taken from a Gaussian distribution, which for the Dicke model has a width of $2/N$ for each degree of freedom, where $N$ is the number of atoms.

Figure \ref{fig:otoc}(b) shows the behavior of the OTOC as a function of time. Note that the exponential growth rate of the OTOC can be attributed to the variance $\sigma^2(t)$ \cite{Lewis2019}, since the growth rate of both quantum and classical variances are the same until saturation \cite{fox1994}. Here, the variance of the electrical coordinate $\sigma^2(\tilde{I}_{L_1})$ as a function of time was calculated by averaging over the ensemble of two hundred fifty-six trajectories. The growth rate $\Lambda$ of the OTOC is obtained by fitting the curve with a monomial function indicated by the black solid line in the same figure. Note that this is the same procedure as the one reported in Ref \cite{chavez2019}. We find that the slope of the linear regression is $\Lambda=0.2865$. At long times, the OTOC saturates and fluctuates around its asymptotic value, due to the finite size of the phase plane. In this regime, the quantum-classical correspondence no longer holds. For comparison purposes, we also calculate the mean Lyapunov exponent by averaging over the resulting Lyapunov exponents of the two hundred fifty-six initial conditions. We obtained a mean Lyapunov exponent of $\hat{\lambda}=0.1147$, which is nearly two times smaller than the grown rate of the OTOC as was predicted in \cite{Lewis2019}. This confirms the  quantitative correspondence between both indicators.

\section{Conclusions}

In summary, we have presented an electrical version of the classical Dicke model, implemented by means of active, synthetic electrical networks. Our setup makes use of two non-linearly coupled active LC oscillators, where the system parameters $\omega$, $\omega_0$, $\gamma$, and initial conditions are independently controlled by voltage signals, thus making the configurations of the system easily accessible, and capable of exploring many remarkable features of the Dicke model. We demonstrated that our experimental setup allows for the direct observation of both phase transitions as well as the exploration of a wide range of excitation energies and Hamiltonian parameters, unveiling the different dynamical regimes, from regular to fully chaotic. Experimental phase planes and Lyapunov exponents were used to study regular and chaotic trajectories. The high resolution of our measuring equipment allowed us to show the existence of a phase transition by following the classical analogous of the mean photon number and the atomic inversion as a function of the coupling parameter. In particular, we showed that the ground state goes from an unexcited normal phase (no photons and no excited atoms) to a symmetry-broken super-radiant phase (field and atomic acquire macroscopic occupations).  Also, we revealed the presence of an excited-state phase transition by studying the stability conditions of the equilibrium point $\mathbb{X}_0$ for small perturbed initial conditions, as a function of the interaction strength. $\mathbb{X}_0$ point goes from a center point ($\gamma<\gamma_c$) to a hyperbolic point ($\gamma\geq\gamma_c$), yielding dramatic alterations in the available phase plane. Interestingly, in our experimental setup, the initial perturbations are produced by the intrinsic electronic noise in the device, which allows us to define a ``classical uncertainty region.'' In all cases, exhaustive numerical simulations were performed to show quantitative and qualitative agreement between theory and experiment. Finally, motivated by the current interest in the OTOC growth rate as a quantum signature of classical chaos, we calculated this quantity for an ensemble of experimental trajectories. Remarkable, the resulting growth rate coincides with the classical Lyapunov exponent, thus demonstrating that, indeed, our non-linear electrical oscillators display all important features of the classical Dicke model. Because of its simplicity, our experimental setup constitutes a remarkable platform for exploring the richness of non-linear systems in terms of chaos and regularity.

\section{Acknowledgements}

M. A. Q. J. and R. J. L. M. were supported by CONACyT under the project CB-2016-01/284372. J. Ch-C and J. G. H. thanks partial financial support from DGAPA-UNAM Grant No. IN109417. M. A. Q. J. and J. L. A. thankfully acknowledge financial support by CONACyT under the project A1-S-8317.

\appendix

\section{Electronic Design of the Classical Dicke Model}

In this appendix we provide a detailed description of the electronic implementation of the classical Dicke model. We start by writing the classical Hamiltonian (\ref{eq:classical}) in terms of electrical variables, considering the map: $(p,q,P,Q)\rightarrow(\tilde{V}_{C_1},\tilde{I}_{L_1},\tilde{V}_{C_2},\tilde{I}_{L_2})$.  Thus, Eq. (\ref{eq:classical}) becomes

\begin{equation}
\begin{aligned}
H&=\frac{\omega}{2}(\tilde{V}_{C_1}^2+\tilde{I}_{L_1}^2)+\frac{\omega_0}{2}(\tilde{V}_{C_2}^2+\tilde{I}_{L_2}^2)\\&+\gamma \tilde{I}_{L_1} \tilde{I}_{L_2}\sqrt{4-\tilde{V}_{C_2}^2-\tilde{I}_{L_2}^2}-\omega_0,
\end{aligned}
\label{eq:hamilosc}
\end{equation}
where $\omega^2=1/L_1C_1$ and $\omega_0^2=1/L_2C_2$ define the natural frequencies for first and second oscillator, respectively. If we set $|L_1|=|C_1|$ and $|L_2|=|C_2|$, then $\omega=1/|L_1|=1/|C_1|$ s$^{-1}$ and $\omega_0=1/|L_2|=1/|C_2|$ s$^{-1}$. To explore the dynamical properties of the Hamiltonian (\ref{eq:hamilosc}), we obtain the associated classical equations of motion, which read

\begin{equation}
\begin{aligned}
C_1\dot{\tilde{V}}_{C_1}&=-\gamma C_1 \tilde{I}_{L_2}\sqrt{4-\tilde{V}_{C_2}^2-\tilde{I}_{L_2}^2}-\tilde{I}_{L_1},\\
L_1\dot{\tilde{I}}_{L_1}&=\tilde{V}_{C_1},\\
C_2\dot{\tilde{V}}_{C_2}&=\frac{\gamma C_2 \tilde{I}_{L_1} \tilde{I}_{L_2}^2}{\sqrt{4-\tilde{V}_{C_2}^2-\tilde{I}_{L_2}^2}}-\gamma C_2 \tilde{I}_{L_1} \sqrt{4-\tilde{V}_{C_2}^2-\tilde{I}_{L_2}^2}-\tilde{I}_{L_2},\\
L_2\dot{\tilde{I}}_{L_2}&=\tilde{V}_{C_2}-\frac{\gamma L_2 \tilde{V}_{C_2} \tilde{I}_{L_1} \tilde{I}_{L_2}}{\sqrt{4-\tilde{V}_{C_2}^2-\tilde{I}_{L_2}^2}}.
\end{aligned}
\label{eq:motion}
\end{equation}

Our setup reproduces the dynamics of the motion equations described by Eq. (\ref{eq:motion}) (which effectively corresponds to a pair of non-linearly coupled LC oscillators) using active electrical networks of operational amplifiers (OPAMPs). These networks consist of interconnecting resistors, capacitors and OPAMPs to conform basic electrical networks where their voltage transfer functions are analogous to the mathematical operations. Thus, with few passive components around the operational amplifiers, it is possible to perform operations of addition, subtraction, integration, and amplification. Although these type of networks are commonly used to perform linear operations, a large number of active components can also implement nonlinear operations, such as logarithms, antilogarithms and products between variables. A drawback of nonlinear configurations is their high temperature-dependence which can lead to functional problems \cite{ruben2013}. A possible solution to implement nonlinear function consists of employing integrated analog multipliers. In fact,   analog multipliers AD633JN (four-quadrant voltage multipliers chips with a typical error less than 1\%) were used to implement the nonlinear terms of our model, such as squaring, square rooting, multiplication, and division between variables.

\begin{figure}[t!]
\centering
\includegraphics[width=8cm]{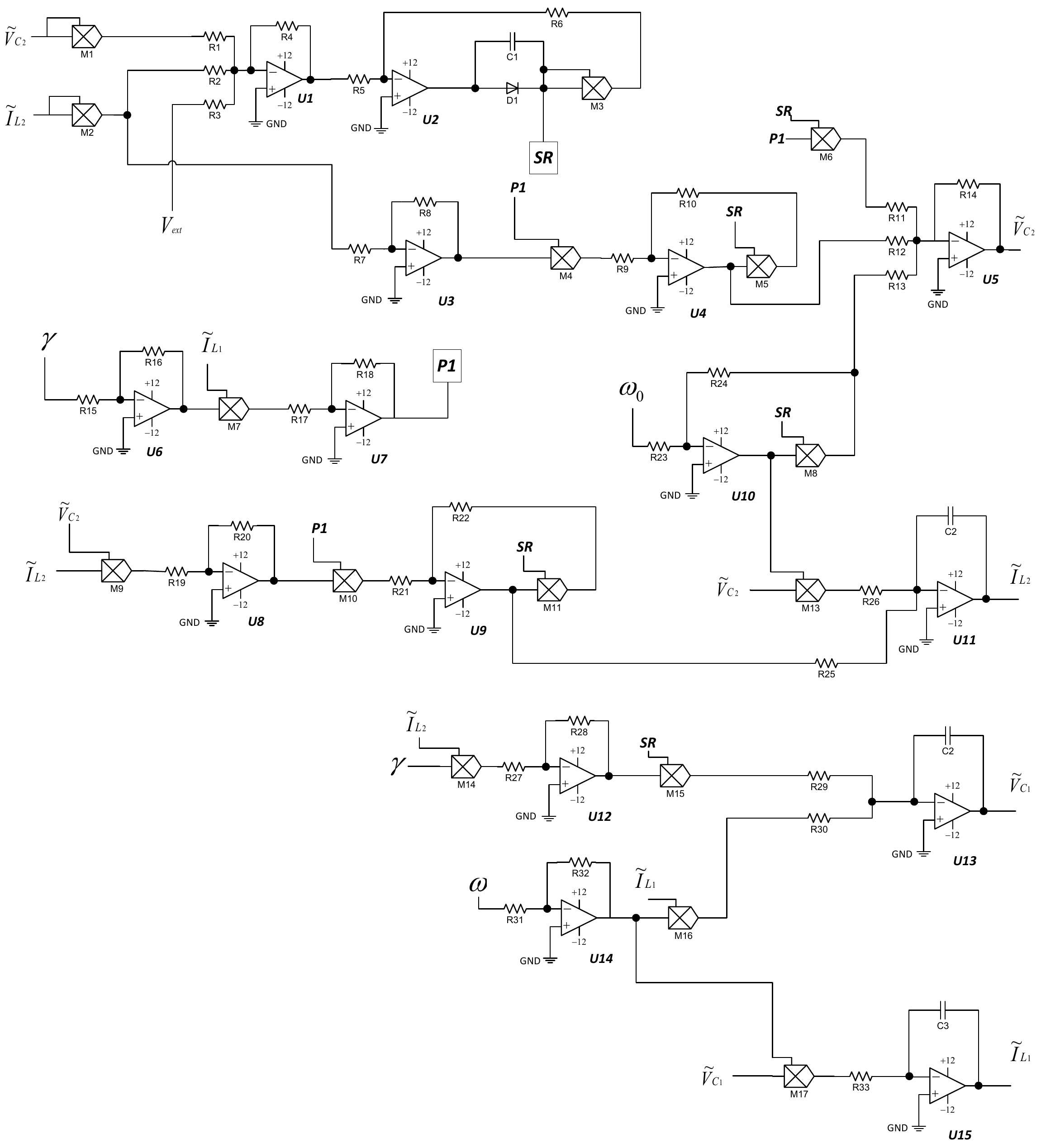}
\caption{Scheme of the electronic circuit of two nonlinearly coupled LC oscillators, described by Eq. (\ref{eq:motion}). The values for the resistors and capacitors are given in the text. The labels $SR$ and $P1$ refer to interconnection of internal signals; whereas $GND$ refers to the ground.}
\label{fig:circuitscheme}
\end{figure}

The electronic implementation of Eq. (\ref{eq:motion}) presents two difficulties: 1) some variables exhibit amplitudes that exceed power supply limits and 2) the system parameters $(\omega,\omega_0, \gamma)$, and initial conditions $(\tilde{V}_{C_1},\tilde{V}_{C_2},\tilde{I}_{L_1},\tilde{I}_{L_2})$ should be controlled in an accessible way. To solve the first problem, we adopt scaling factors in gain amplifier configurations in order to obtain a half-scale condition. Particularly, the analog multipliers normalize the input signals by a factor of about 10. For the second problem, we introduce system parameters, and initial conditions, via voltage signals provided by digital-to-analog converters (MCP4921, resolution: 12 bits) which are communicated to a master 8-bits microcontroller (PIC18 family) by serial peripheral interface (SPI) protocol. Each voltage value can be individually configured via software. With this configuration we avoid variations in the properties of the electronic components of the device. The schematic of the electronic circuit is shown in Fig. \ref{fig:circuitscheme}. There, $R_j$ , $C_j$ , $U_j$ and $M_1$ stand for resistors and capacitors, general-purpose operational amplifiers MC1458, and analog multiplier AD633JN, respectively. $D_1$ denotes a fast recovery diode 1N4148. The resistor and capacitor values used in the circuit are the following: $R_1=R_2=R_3=R_4=R_7=R_{17}=R_{19}=R_{27}=1K\Omega$, $R_5=R_6=R_8=R_9=R_{10}=R_{11}=R_{13}=R_{16}=R_{18}=R_{20}=R_{21}=R_{22}=R_{24}=R_{25}=R_{26}=R_{28}=R_{29}=R_{30}=R_{32}=R_{33}=10K\Omega$, $R_{12}=100K\Omega$, $R_{15}=R_{23}=R_{31}=5K\Omega$  $C_1=0.01\mu$F  and $C_2=C_3=C_4=C_5=10\mu$F. The labels $SR$ and $P1$ refer to interconnection of internal signals.

\begin{figure}[h]
\centering
\includegraphics[width=8 cm]{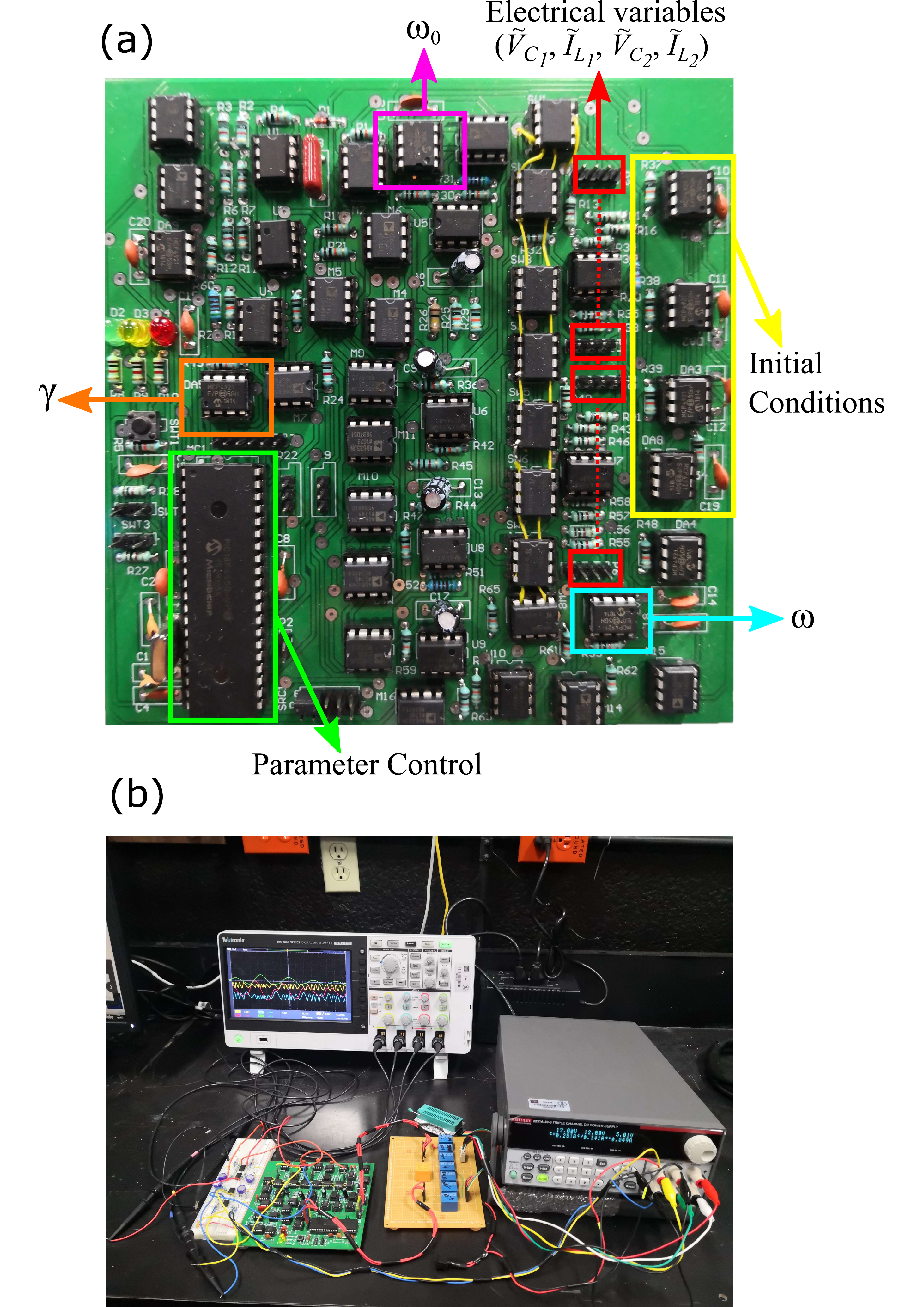}
\caption{(a) Printed circuit board of the classical Dicke model. (b) Complete experimental setup.}
\label{fig:circuit}
\end{figure}

Electronic components were mounted and soldered on a printed circuit board (PCB, 15x15 cm) to avoid faulty contacts and poor stability. Figure \ref{fig:circuit}(a) shows the electronic circuit realization of the system (\ref{eq:motion}) on the PCB. In the same figure, the parameters of the system are signaled with orange, cyan and magenta squares, and the initial condition module with yellow; all of them governed by a microcontroller (green square) via software. The red square indicates the output electrical variables $(\tilde{V}_{C_1},\tilde{I}_{L_1},\tilde{V}_{C_2},\tilde{I}_{L_2})$. The whole experimental setup mounted in the laboratory is shown in Fig. \ref{fig:circuit}(b).

\begin{figure}[t!]
\centering
\includegraphics[width=\linewidth]{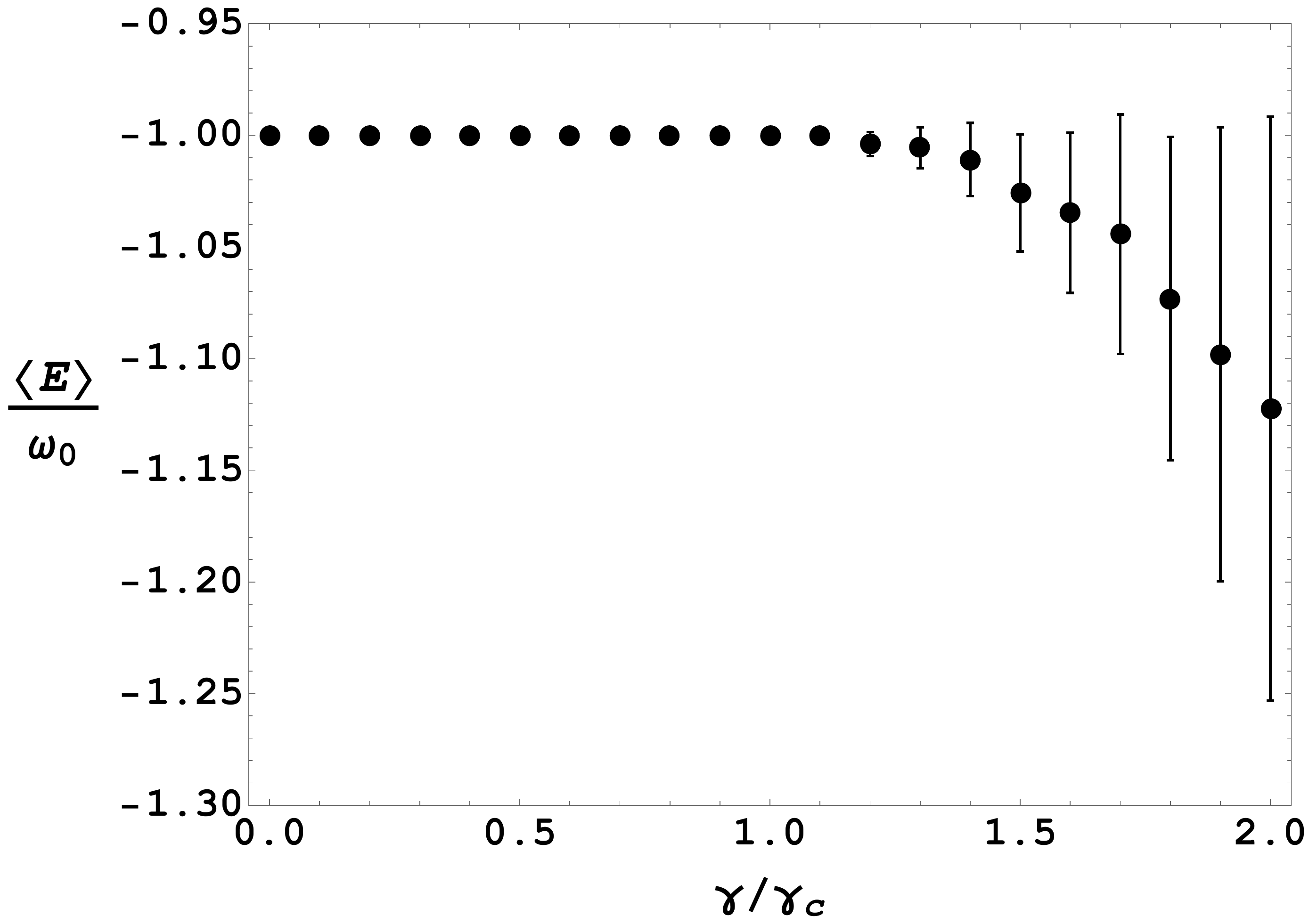}
\caption{Expectation energy value $\left< E \right> $ of the classical Dicke Hamiltonian as a function the coupling strength. Error bars correspond to the standard deviations.}
\label{fig:energytime}
\end{figure}

In spite of the robustness, flexibility and good agreement with theoretical results, the proposed circuit is not ideal, it presents a slight energy loss which becomes evident in the super-radiant phase. In Fig. \ref{fig:energytime}, we plot the expectation energy value $\left< E \right> $ as a function the coupling strength.  For this plot, experiments were carried out in the resonance case $(\omega=\omega_0=1)$.  The standard deviations are shown by error bars. It is clear that if one evaluates the Hamiltonian (\ref{eq:hamilosc}) at the initial condition $(\tilde{V}_{C_1},\tilde{I}_{L_1},\tilde{V}_{C_2},\tilde{I}_{L_2})=(0,0,0,0)$, the expectation energy value should be always $E=-1$ irrespective of $\gamma$. Despite this, Fig. \ref{fig:energytime} shows a decay in the expectation energy value $\left< E \right> $. Notably, the losses only gain relevance in the super-radiant phase, that is, for coupling values larger than the critical value $\gamma_c$. In addition, note that as $\gamma$ increases, the standard deviations in $\left< E \right> $ so do, this is because the accessible phase space grows with $\gamma$, causing that the trajectories evolve freely.

We find that the energy dissipation in the circuit does not come from a simple Joule-heating loss effect. The electrical network presents highly complex energy exchange, mainly due to the operational amplifiers energy exchange with the power source which polarizes them. Remarkably, the dynamical behavior of the system is essentially unaffected by the dissipation effect. Furthermore, our capacity to control parameters and initial conditions is not seen limited by such phenomenon.

\section{Lyapunov Exponent Calculation}

A notable property of a chaotic system is its high sensitivity to initial conditions,  a fact that leads to the concept of the Lyapunov exponent, which has been used as an indicator of chaos for a long time. It measures the average exponential divergence of conditions infinitesimally nearby to the initial state. In general terms, there are as many Lyapunov exponents as the dimension of the phase plane. However, in many
applications, it is sufficient to calculate only the largest Lyapunov exponent. A system with a positive largest Lyapunov exponent is defined to be chaotic \cite{liapunov, oseledets1968, benettin1980}.

In the subsection ``Regularity and chaos", we estimated the largest Lyapunov exponent of each point of Fig \ref{fig:mapLyapunov} (a), (b), (c.1), (c.2), (c.3), (c.4) and (d), using the experimental time series, that is, the temporal evolution of the variables $(\tilde{V}_{C_1},\tilde{I}_{L_1},\tilde{V}_{C_2},\tilde{I}_{L_2})$, following the proposed method by Ref. \cite{rosenstein1993} which is fast and easy to implement. Below we summarize the algorithm.

\textbf{\textit{Step 1}}. Reconstruction of the attractor dynamics from a single time series. In our case, all state variables of the system are available, namely, $(\tilde{V}_{C_1},\tilde{I}_{L_1},\tilde{V}_{C_2},\tilde{I}_{L_2})$. The trajectory $\mathbb{X}$ of the system can be expressed as:

\begin{equation}
\mathbb{X}=\left\{x_1, x_2,...,x_M\right\},
\end{equation}
where $x_i=[\tilde{V}_{C_1}(t_i),\tilde{I}_{L_1}(t_i),\tilde{V}_{C_2}(t_i),\tilde{I}_{L_2}(t_i)]$ is the state of the system at a discrete time $t_i$. $M$ is the number of data points in the time series.

\textbf{\textit{Step 2}}. Calculate the nearest neighbor of each point on the trajectory. The nearest neighbor, $x_{\hat{i}}$, minimizes the Euclidean distance to the reference point $x_i$, that is,

\begin{equation}
d_i=min \parallel x_i-x_{\hat{i}} \parallel .
\end{equation}
There, the nearest neighbors should have a temporal
separation greater than the mean period $\overline{T}$ of the time series, $\parallel x_i-x_{\hat{i}} \parallel > \overline{T}$.

\textbf{\textit{Step 3}}. The largest Lyapunov exponent is calculated using a least-squares fit to the average separation of neighbors defined by

\begin{equation}
y(t_i)\approx \frac{1}{t_s} \left< \ln d_j(t_i) \right>,
\end{equation}
where $t_s$ is the sampling period and $\left< . \right>$ denotes the average over all values of $j$.

Exhaustive numerical simulations were performed by varying the number of data points of experimental time series, which were smoothed using the moving average filtering method \cite{smith2013}. From our results, we find that Lyapunov exponents with values close to theoretical values were obtained with a average time-window size equal to 750 ms, and a time series of 120 s.

\bibliography{apssamp}

\end{document}